\documentclass[]{pasj01}
\draft
\usepackage{url}
\usepackage{threeparttable}
\usepackage{multirow}

\begin{document}
\Received{}%{yyyy/mm/dd}
\Accepted{}%{yyyy/mm/dd}
%\Published{yyyy/mm/dd}

\title{Two twin binaries with nearly identical components: KIC 4826439 and KIC 6045264}

%%% begin:list of authors
% Do NOT capitalize all letters in "textsc".
\author{Jia \textsc{Zhang}\altaffilmark{1,2}}%
\altaffiltext{1}{Yunnan Observatories, Chinese Academy of Sciences, Kunming 650216, China}
\altaffiltext{2}{Key Laboratory of the Structure and Evolution of Celestial Objects, Chinese Academy of Sciences, Kunming 650216, China}
\email{zhangjia@ynao.ac.cn}

\author{Sheng-Bang \textsc{Qian},\altaffilmark{1,2}}

\author{Shu-Min \textsc{Wang}\altaffilmark{3}}
\altaffiltext{3}{School of Mathematics and Physics, Handan University, Handan 056005, China}

\author{Yue \textsc{Wu}\altaffilmark{4}}
\altaffiltext{4}{Key Laboratory of Optical Astronomy, National Astronomical Observatories, Chinese Academy of Sciences, Beijing 100012, China}

\author{Lin-Qiao \textsc{Jiang}\altaffilmark{5}\thanks{School of Physics and Electronic Engineering, Sichuan University of Science \& Engineering, Zigong 643000, China}}
\altaffiltext{5}{School of Physics and Electronic Engineering, Sichuan University of Science \& Engineering, Zigong 643000, China}
\email{linqiao@ynao.ac.cn}

%%% end:list of authors

%% `\KeyWords{}' always has to be placed before `\maketitle'.
 %Do NOT move this preamble from here!

\maketitle

\begin{abstract}
Two twin binaries, KIC 4826439 and KIC 6045264, with very similar component stars were found photometrically based on \textit{Kepler} eclipsing binary light curves. The absolute parameters of the massive components are 1.156(0.03)$M_\odot$, 1.881(0.02)$R_\odot$, 6065K for KIC 4826439, and 0.874(0.3)$M_\odot$, 1.206(0.02)$R_\odot$, 6169(30)K for KIC 6045264. The differences between the components are less than two percents for all the parameters. A very low proportion of the twin binaries ($2/1592\approx0.13\%$) was found, which does not support the previous findings of the excesses of twins on binary mass ratio distribution, but support a deficiently low proportion of twins. A new method is practiced to work out the absolute parameters of the two twins without the radial velocities. This method requires the solution of the light curves, the spectra and the evolutionary isochrones of covering the complete stellar parameter space, simultaneously. We also studied their evolution tracks that: KIC 4826439 will experience an unstable mass transfer stage followed by an unclear ending, and KIC 6045264 will become a single star via an over-contact phase. It seems highly unlikely that the two twin binaries will produce twin degenerate binaries, although they have quite similar components.
\end{abstract}

\section{Introduction}

Binaries with similar components, often characterized quantitatively by mass ratio $q\simeq1$, are effectively called twin binaries. These kind of binaries were first noted by \citet{1979AJ.....84..401L} who found a narrow peak at $q\simeq0.97$ on the distribution of mass ratios of double-lined spectroscopic binaries. Afterwards, a high twin population (20\% - 25\%) for massive binaries are proposed based on the double-lined spectroscopic binaries in the Small Magellanic Cloud \citep{2006ApJ...639L..67P}. The other similar twin populations are also presented \citep{2000A&A...360..997T,2003A&A...397..159H,2006A&A...457..629L,2009AJ....137.3442S}. However these high twin populations are argued by \citet{2014MNRAS.445.2028C} for the strong selection effects on spectral observations, that the double-lined binaries bias towards twins binaries. If the binaries are single-lined, the binaries are surely not twins.

Besides the observations on twin binaries, the theoretical works were also carried out to draw the twins population in binaries. \citet{2000MNRAS.314...33B} predicted that there should be a continuous trend for closer binaries to equalize the components masses, by accreting material from circumbinary molecular cloud to the original small `seed' protobinary system. While this model requires that the original protobinary seed is initially very small and accrete a large amount of mass, typically 10 times its original mass, to produce twin binaries. Nevertheless, the small `seed' is difficult to form due to the radiation feedback \citep{2006ApJ...641L..45K,2007ApJ...656..959K}, therefore \citet{2007ApJ...661.1034K} propose another formation mechanism for twin binaries. During the pre-main-sequence evolution, a phase of deuterium shell burning will swell the protostars to tenths of an AU in radius, and will make close binaries overflow their Roche lobes to transfer mass from more massive to the smaller component, thus to drive the mass ratio toward equality on a dynamical timescale \citep{2007ApJ...661.1034K}. This mechanism is natural and inevitable for close binaries. Conversely, \citet{2012Sci...337..444S} studied the mass transfer between massive stars theoretically and find no preference for equal mass binaries.

Most of the twin binaries were found by the spectroscopic data. Consequently, an attempt of finding photometric twin binaries is carried out, based on the large database of \textit{Kepler} binary light curves from the \textit{Kepler} spacecraft telescope \citep{2010Sci...327..977B,2010ApJ...713L..79K,2010ApJ...713L..92C}. The process for searching twins is described in the next section. Then the two new discovered twin binaries, KIC 4826439 and KIC 6045264, will be studied in detail in the latter sections, where the absolute parameters were calculated by a new method and the evolutionary endings were computed. We summarize our conclusions in the last Section \ref{sec:sum_dis}.

\section{Searching twins photometrically}

\subsection{selecting the initial candidates}

The catalog of \textit{Kepler} Eclipsing Binaries \citep{2011AJ....141...83P,2011AJ....142..160S,2012AJ....143..123M,2014AJ....147...45C}, containing about 2600 objects, is our sample library for searching twin binaries. The most important feature of \textit{Kepler} observations is the generally uninterrupted brightness monitoring in a 115 $deg^{2}$ fixed field of view, in where the telescope can find almost all the eclipsing binaries with period of less than 4 years, or 1 year if more reliable. Moreover, the high limiting magnitude (21st magnitude) and the high photometric precision (20 per million) make the telescope can detect small light variation binaries, which is important especially for contact binaries in small inclinations. Most of the Kepler eclipsing binaries are bright which is less than 16th magnitude in \textit{Kepler} band (close to $R_c$ band). There could be selection effects on transiting planet objects \citep{2013ApJ...762...41G} due to the mission goal, but should be no selection effect on binary objects. Because of these features, the \textit{Kepler} Eclipsing Binaries are a good library for searching twin binaries, as well as for investigating the proportion of twins. It is believed that all the mass ratios can be equally discovered, from $\sim0$ to $\sim1$ (Note that \textit{Kepler} had discovered many planets revolving around stars, which can be considered, technically, as binaries with tiny mass ratios.)

Our method of searching twins is very simple. If a binary has very similar component stars (in masses, radiuses and temperatures etc.), its light curves should have very similar primary and secondary eclipses (only for circular orbits, explained below). Therefore, the binary light curves with similar eclipses are selected to be the initial twin candidates. The catalog of Kepler Eclipsing Binaries\footnote{downloaded from http://keplerebs.villanova.edu/search} provided the eclipsing depthes and widthes for almost every binary object. We compared the primary and secondary eclipses on depth and width for every binary, and selected 26 binaries with quite similar eclipses (the differences on depthes and widthes are all smaller than 2 percent) as the initial candidates. They are KIC 202083222, KIC 11091082, KIC 7818448, KIC 4037163, KIC 5077994, KIC 6045264, KIC 8608310, KIC 9402652, KIC 8444552, KIC 5598639, KIC 9953894, KIC 4550750, KIC 9912977, KIC 8330575, KIC 8244173, KIC 5903301, KIC 4826439, KIC 5565486, KIC 3449540, KIC 4365461, KIC 6891512, KIC 4758368, KIC 10480952, KIC 8302455, KIC 4157488 and KIC 5988465.

\subsection{calculating the mass ratios}

The binaries with mass ratio greater than 0.95 are defined as twins in this paper. In order to find them, the technique of mass ratio search was used. The mass ratio search, often called q-search, is a commonly used technique to find the binary mass ratio from the light curves. The specific steps are: First, fix the mass ratio at a specific value, and then fit the light curves with other parameters adjustable. Next, change the fixed value of the mass ratio, and repeat the fitting work. Generally, the mass ratio will be fixed at many values that can cover a large enough range. After the fittings on all the mass ratios, the fitting residuals were plotted against the mass ratios which usually form a dip. Then the mass ratio with the lowest fitting residual (corresponding to the best fitting and so the best mass ratio) is thought to be close to the real value. The final mass ratio can be obtained when itself is set to be adjustable using the best mass ratio as the initial guess value.

The q-search was carried out at mass ratio around 1 on the 26 initial candidates. If the fitting residual - mass ratio curve (q-search curve) for a candidate show a dip with minimum near 1 (see the upper panels of Figure \ref{fig:wd} for example), the candidate is a twin binary. However, if the q-search curve change monotonically around 1, indicating the mass ratio is far from 1, then the candidate is not a twin binary. The results show that only two twins, KIC 4826439 and KIC 6045264, were found, see the upper panels of Figure \ref{fig:wd}.

Among the 26 initial candidates, the two twin binaries have the most similar primary and secondary eclipses. This indicates that the similarity of the eclipses is a good index for searching twins; and also indicates that the other binaries, with larger differences on eclipses, are very unlikely to be twins. The similarity of eclipses can be seen from the lower panels of Figure \ref{fig:wd}. Besides the two twins, there are four candidates found with mass ratio $q$ relatively close to 1, they are KIC 4758368 ($q=1.26$), KIC 5565486 ($q=0.86$), KIC 9402652 ($q=0.92$) and KIC 9912977 ($q=1.2$). For the candidates with mass ratio far away from 1, we did not compute their specific values.

There are two points need to be explained. First, only the binaries with period less than 5 days in the catalog of \textit{Kepler} Eclipsing Binaries were used to search twins, because it is difficult to obtain reliable mass ratios for the long period binaries or eccentric orbit binaries based purely on the single color \textit{Kepler} binary light curves. Second, the light curves used in the q-search work were not the entirely original data in BJD form, but the folded and binned data in phase form (before the folding and binning, the data were normalized first which will be described in Section \ref{sec:data_normal}). It is much easier to get a smooth q-search curve using the binning phase data, because the binning data can significantly remove the asymmetry on the binary light curves caused by magnetic activities, pulsation, etc., and also significantly reduced the amount of data. It should be noted that the binning data points are evenly distributed along the curve (the distances between the adjacent points along the curve are roughly the same), not along the phases (the phase intervals between the adjacent points are the same). It is believed that this kind of points is more suitable for the analysis, because more weight is put on the important eclipsing parts.

\subsection{the proportion of twins}

With the restriction of period less than 5 days, there are 1592 eclipsing binaries, and so the proportion of twin binaries (mass ratio $q$ greater than 0.95) is $2/1592\approx0.13\%$, which is much smaller than the uniform distribution of 5\%.

There are some factors may reduce our searching results: 1, there may be eccentric orbital binaries with period of less than 5 days, but our method can not find twin binaries with eccentric orbits. However, this situation is extremely rare if there is, because the eccentric binaries with period less than 5 days are very rare. 2, The \textit{Kepler} Eclipsing Binaries may missed or wrongly calculated the widthes and depthes for some binaries, in where there maybe twins. This situation is also very rare. 3, Some other light variations on the binary light curves, such as magnetic actives and pulsations, may interfere with the q-search works, that may lead to miss some twins. This is an important factor, because the mass ratio derived from the light curve analysis were often affected by the non-binary light variations heavily. So the proportion of twins may be underestimated in some degree.

Despite the factors mentioned above, the real proportion of twins is highly unlikely to be comparable to the uniform distribution 5\%, because the discovered proportion is about 40 times lower, i.e. only $2/1592\approx0.13\%$. Even supposing the 26 initial candidates are all twins, the proportion is just $26/1592\approx1.6\%$ which is still considerable lower than 5\%, and this is almost impossible. Therefore, our search result does not support the excesses on twin distribution but indicates the contrary conclusion that the proportion of twins is very low.

\section{Data processing and observation}

The bases of our works include the binary light curves from the \textit{Kepler} mission that can be obtained from Mikulski Archive for Space Telescopes (MAST)\footnote{\url{http://archive.stsci.edu/}}, and the spectral observations that were carried out on May 27th 2016 with the 2.4 m telescope at Yunnan Observatories. In addition to the \textit{Kepler} photometric data, the \textit{Kepler} Eclipsing Binary Catalog \citep{2011AJ....141...83P,2011AJ....142..160S,2012AJ....143..123M,2014AJ....147...45C} is also the essential basis and guidance. The catalog provides the key parameters of binaries including orbital periods, eclipse widthes and depthes, which are very helpful to our search works.

\subsection{Kepler light curves normalization} \label{sec:data_normal}

The \textit{Kepler} photometric fluxes we choosed are the Pre-Search Data Conditioning (PDC) fluxes whose purpose is `to identify and correct flux discontinuities that cannot be attributed to known spacecraft or data anomalies' (Kepler Data Processing Handbook\footnote{\url{http://archive.stsci.edu/kepler/manuals/KSCI-19081-001_Data_Processing_Handbook.pdf}}; \citet{Fraquelli2014}). In spite of this, there are still many unreal jumps and long term variations on the light curves, and they need to be eliminated before the light curve analysis. Typically, almost all the jumps occurred at the gaps on the light curves, about half of them are the borders of observation quarters of the \textit{Kepler} satellite, and the other half are the missing points inside the quarters. Therefore, the flux normalization performed on each segment of successive data points will naturally eliminate the jumps.

For the unreal long term (weeks or longer) variation, we can not separate them from the real variations. Fortunately, the binary geometric variations are always short term, no matter what the orbital period is\footnote{Even for the long period binaries, the light variations only take place at the eclipses, which is still short.}. So the elimination of the long term variations, containing both real and unreal composition, don't affect the analyses on the binary geometric light variations.

The steps of the light curve normalization are as follows: First, the light curves of each object are divided into a number of segments according to the gaps on the light curves. Each gap with width bigger than a given time interval\footnote{In this paper, the time interval is set to be one day arbitrarily. The gaps between the observation quarters are always larger than one day.} will be treated as a break. Then the light curves will split at all the breaks. Second, the light curve of each segment is fitted by a cubic polynomial model, and then the light curves fluxes $F_{obs}$ are divided by their fitting values $F_{fit}$ to get the normalized fluxes $F_{nor}$, i.e. $F_{obs}/F_{fit}=F_{nor}$. During the fitting, the large scatter points with residuals larger than three times the standard deviations are removed, and the light curves will be refitted. If new large scatter points appear, they are removed again before another fitting until no large scatter point appears, or until the fitting times reaches a given number (here ten). The large scatter points not only include the bad points on the light curves, but also include the real points such as sudden deep eclipse points or magnetic flare points. Thus, the fitting curves are representative for the long term variations. Third, all the normalized separated segments of light curves are stitched together to get a whole normalized light curve which is flat on long scale and without jump.

For the further analysis on the two twins, the original BJD form data, instead of the binning phase data in the q-search work, were used, because the period change rate can be obtained on BJD data. However, not all the original data were used in the analysis. There are 64,783 and 64,787 data points for KIC 4826439 and KIC 6045264 respectively, which are too large to runnable for the light curve model. So only a part of them, 11,000 points (KIC 4826439) and 8,400 points (KIC 6045264), were selected randomly from the original data points and they are listed in Table \ref{tab:lcs}. In the selection, the points around eclipses were more likely to be selected. Specifically, if the selected points are folded into phase, 7,200 (KIC 4826439) and 5,100 (KIC 6045264) points will locate within the eclipses (primary or secondary) with a width of 0.12 (KIC 4826439) and 0.17 (KIC 6045264). It is equivalent to say that the points around eclipses are six (KIC 4826439) and three (KIC 6045264) times denser than the other parts. This way of reduction is the same as we did in the binning phase data, because we thought that the important eclipsing parts should be emphasised with more data in the analysis work.

\begin{table}[h!]
  \tbl{The light curves of KIC 4826439 and KIC 6045264.}{
  \begin{tabular}{cc|cc} %{cc|cc|cc|cc|cc|cc}
      \hline
      \hline
      \multicolumn{2}{c|}{KIC 4826439} & \multicolumn{2}{c}{KIC 6045246} \\
      \hline
      BJD-2454833 & mag & BJD-2454833 & mag  \\
      \hline
      131.51270 &  0.153333 & 131.88021 & -0.058955   \\
      131.77835 & -0.000984 & 131.98238 & -0.072583   \\
      131.88052 & -0.003914 & 132.02325 & -0.071540   \\
      ...&...&...&...  \\
      \multicolumn{2}{c|}{11000 points} & \multicolumn{2}{c}{8400 points} \\
      ...&...&...&...  \\
      1590.38860 & -0.001903 & 1590.94004 & -0.070302 \\
      1590.96077 &  0.004641 & 1590.96047 & -0.067609 \\
      \hline
      \hline
  \end{tabular}}\label{tab:lcs}
\begin{tabnote}
This table is available in its entirety in machine-readable and Virtual Observatory (VO) forms in the online journal. A portion is shown here for guidance regarding its form and content.
\end{tabnote}
\end{table}

\subsection{the spectroscopic observations and reduction}

\begin{figure}[h!]
\centering
\includegraphics[width=.50\textwidth]{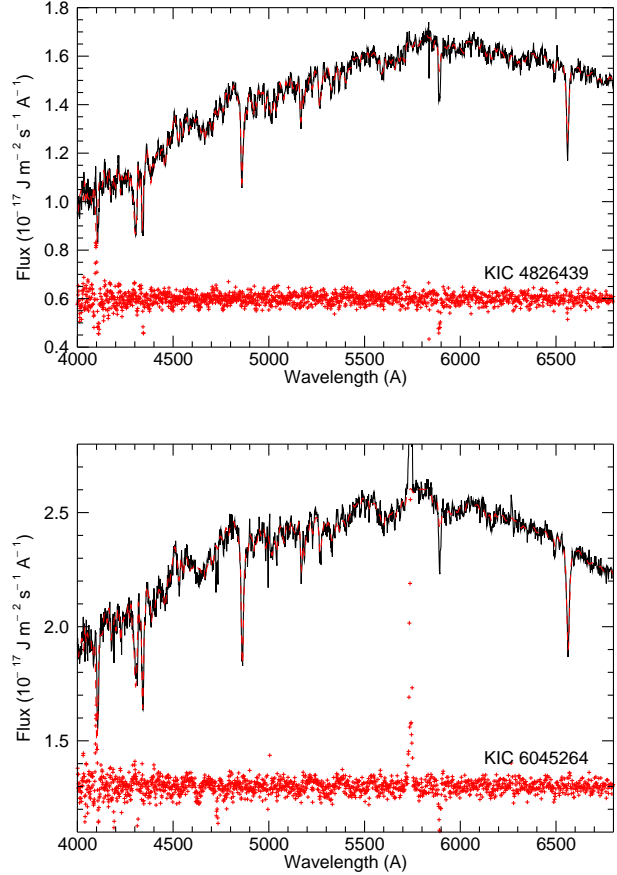}
\caption{The spectra of KIC 4826439 (upper) and KIC 6045264 (lower). The black lines are the observational points, and the red dashed lines are the best fitting models. The horizontal red points on the lower parts are the fitting residuals, and the points of jumping up around 5743 {\AA} (lower panel) are caused by the emission line (shown partly) which is not fitted by the model. \label{fig:spectra}}
\end{figure}

The objects KIC 4826439 and KIC 6045264 were observed spectroscopically using the Lijiang 2.4 m telescope at Yunnan Observatories on May 27th 2016. The terminal YFOSC (Yunnan Faint Object Spectrograph and Camera) was used with a grism of $\sim2000$ resolution. Three spectra were obtained for each object with exposure 900 and 600 seconds. The observations were processed by \textsc{IRAF} (Image Reduction and Analysis Facility) with flat correction and absolute flux calibration. The processed data were tabulated in Table \ref{tab:spectral_data} and will be analysed in Section \ref{sec:The_spectra_analysis}. One spectrum for each object was presented in Figure \ref{fig:spectra}.

\begin{table}[h!]
  \tbl{The spectra of KIC 4826439 and KIC 6045264.}{
  \begin{tabular}{cc|cc|cc|cc|cc|cc} %{cc|cc|cc|cc|cc|cc}
      \hline
      \hline
      \multicolumn{6}{c|}{KIC 4826439} & \multicolumn{6}{c}{KIC 6045246} \\
      \hline
      \multicolumn{2}{c|}{BJD: 2457536.2934} & \multicolumn{2}{c|}{BJD: 2457536.3063} & \multicolumn{2}{c|}{BJD: 2457536.3161} & \multicolumn{2}{c|}{BJD: 2457536.3311} & \multicolumn{2}{c|}{BJD: 2457536.3428} & \multicolumn{2}{c}{BJD: 2457536.3524}  \\
      \hline
      Wavelength & Flux & Wavelength & Flux & Wavelength & Flux & Wavelength & Flux & Wavelength & Flux & Wavelength & Flux  \\
      {({\AA})} & ($\frac{10^{-17}j}{{m^2}{s}\textrm{{\AA}}}$) & ({\AA}) & ($\frac{10^{-17}j}{{m^2}{s}\textrm{{\AA}}}$) & ({\AA}) & ($\frac{10^{-17}j}{{m^2}{s}\textrm{{\AA}}}$) & ({\AA}) & ($\frac{10^{-17}j}{{m^2}{s}\textrm{{\AA}}}$) & ({\AA}) & ($\frac{10^{-17}j}{{m^2}{s}\textrm{{\AA}}}$) & ({\AA}) &  ($\frac{10^{-17}j}{{m^2}{s}\textrm{{\AA}}}$) \\
      \hline
      3600.46 & 0.4989 & 3599.74 & 0.7801 & 3600.34 & 0.7806 & 3600.36 & 1.1175 & 3600.59 & 1.2101 & 3599.67 & 1.1403 \\
      3601.68 & 0.4811 & 3600.94 & 0.4910 & 3601.60 & 0.6741 & 3601.67 & 1.2487 & 3601.89 & 1.1489 & 3601.01 & 1.1367 \\
      3602.89 & 0.4508 & 3602.14 & 0.5297 & 3602.86 & 0.5472 & 3602.99 & 1.2359 & 3603.20 & 1.1454 & 3602.36 & 1.0813 \\
      ...&...&...&...&...&...&...&...&...&...&...&...  \\
      \multicolumn{2}{c|}{2264 points} & \multicolumn{2}{c|}{2267 points} & \multicolumn{2}{c|}{2257 points} & \multicolumn{2}{c|}{2249 points} & \multicolumn{2}{c|}{2250 points} & \multicolumn{2}{c}{2244 points}  \\
      ...&...&...&...&...&...&...&...&...&...&...&...  \\
      7498.57 & 0.8428 & 7498.58 & 1.0149 & 7498.57 & 0.9780 & 7498.67 & 1.4886 & 7498.66 & 1.4846 & 7498.70 & 1.4273 \\
      7500.64 & 0.8485 & 7500.65 & 0.9773 & 7500.64 & 0.9644 & 7500.74 & 1.4773 & 7500.74 & 1.4900 & 7500.78 & 1.4209 \\
      \hline
      \hline
  \end{tabular}}\label{tab:spectral_data}
\begin{tabnote}
This table is available in its entirety in machine-readable and Virtual Observatory (VO) forms in the online journal. A portion is shown here for guidance regarding its form and content.
\end{tabnote}
\end{table}

\section{The light curve analysis of KIC 4826439 and KIC 6045264} \label{sec:lc_analysis}

\subsection{The background information}

Based on the \textit{Kepler} observations, the object KIC 4826439 was first classified as a binary star system \citep{2011AJ....141...83P} and confirmed by the following classifications \citep{2011AJ....142..160S,2014AJ....147...45C}. However, it was also studied recently as a planetary system based on the same \textit{Kepler} observations \citep{2015ApJ...814..130M}, despite its planetary status was argued by the identification works \citep{2014AJ....147..119C,2015ApJS..217...31M}. In this paper, the object KIC 4826439 was analysed and identified as a stellar binary system.

Before the \textit{Kepler} spacecraft begin to observe, two variability surveys on the ``Kepler Field'' were carried out and KIC 6045264 was found as a binary star system with 0.9093 days period \citep{2004AJ....128.1761H,2009AcA....59...33P}. Afterwards its binarity is verified by the kepler observations \citep{2011AJ....141...83P,2011AJ....142..160S,2014AJ....147...45C}.

KIC 4826439 and KIC 6045264 are both detached binaries with period of 2.474295 days and 0.909310 days respectively, and their temperatures were given by \citet{2014MNRAS.437.3473A}. However, more parameters are needed which is exactly the purpose of this work.

\subsection{The binary light curve analysis}

The kepler binary light curves of the two objects, KIC 4826439 and KIC 6045264, were analyzed using the Wilson-Devinney programme \citep{1971ApJ...166..605W,1979ApJ...234.1054W,1990ApJ...356..613W,2007ApJ...661.1129V,2008ApJ...672..575W,2010ApJ...723.1469W,2012AJ....144...73W}. The primary temperatures are obtained from spectroscopy observations which is $6065\pm45$ K for KIC 4826439 and $6169\pm30$ K for KIC 6045264\footnote{Only one temperature was measured from each binary spectra, and this temperature was assigned to the primary star (the star blocked by its companion at phase 0 or 1). For the two binaries here, their components' temperatures should be very close to each other, so the single measured value can stand for both the primary and secondary star.}. In the iterative fitting process of the programme, the exponents $g$ in the bolometric gravity brightening law and the bolometric albedos $A$ for reflection heating and re-radiation is fixed at $g=0.32$ and $A=0.5$, because the stellar envelopes of the component stars are probably convective due to their low temperatures.
The logarithmic law for limb darkening is used, and the coefficients are internally calculated in the programme as a function of temperature $T_{eff}$, surface gravity $\lg g$ and Metallicity $[M/H]$ based on the \citet{1993AJ....106.2096V} table. The temperature $T_{eff}$, surface gravity $\lg g$ and Metallicity $[M/H]$ are also derived from spectroscopy observations. However the effects of $\lg g$ and $[M/H]$ on the light curves are quite small, and they hardly affect the other output parameters.

The exposure time of the \textit{Kepler} long-cadence data is about 30 minutes for each point, and this will cause the smearing effects on the light curves, especially on the eclipse part where the light variation is fast. This smearing effects are dealt with an integration via Gaussian quadrature that is already integrated into the Wilson-Devinney programme \citep{2012AJ....144...73W}, and the number of Gaussian abscissas is set to be 3 in the analysis.

\begin{figure*}[h!]
\begin{minipage}{160mm}
\centering
\includegraphics[width=.45\textwidth]{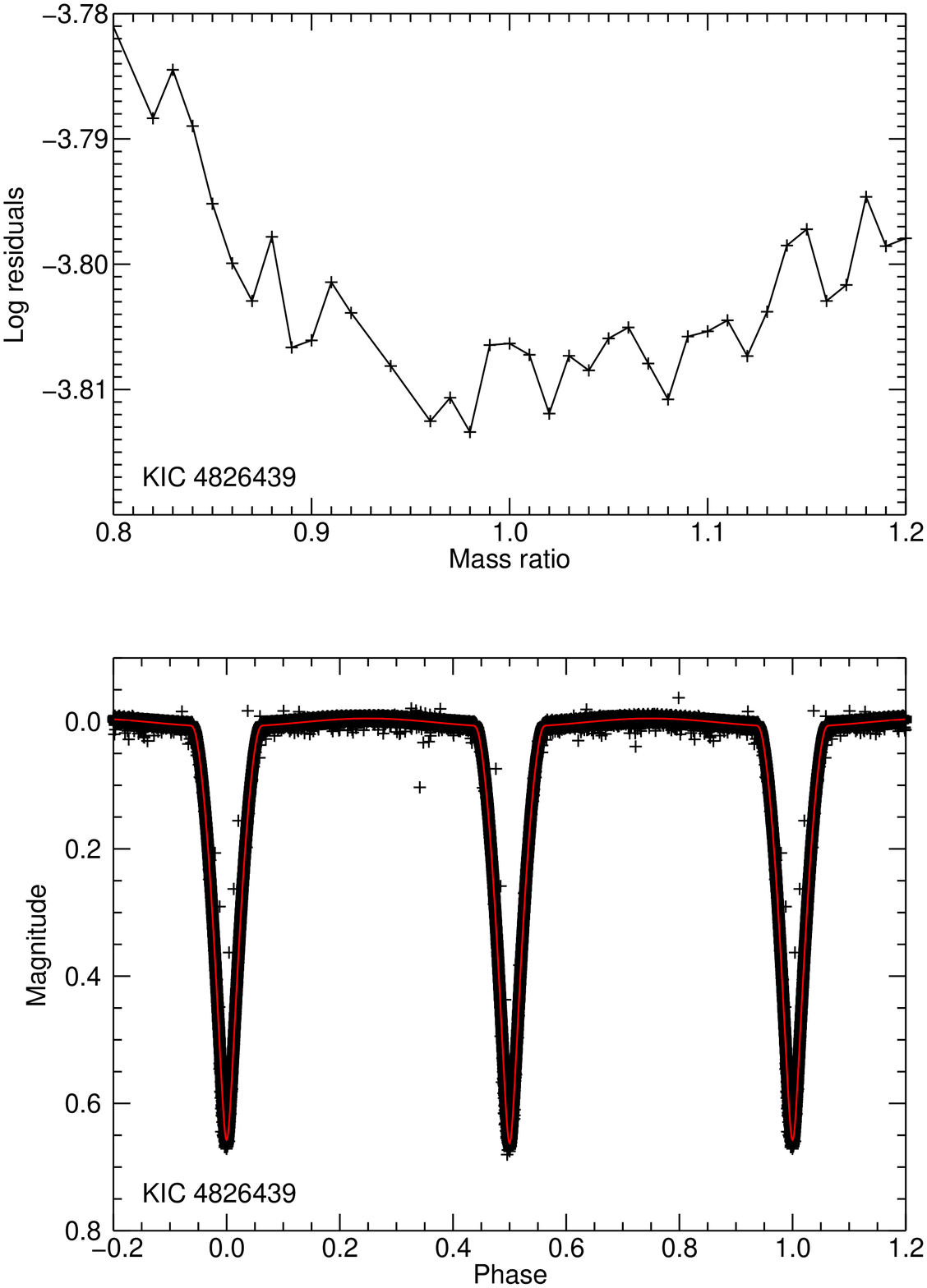}
\includegraphics[width=.45\textwidth]{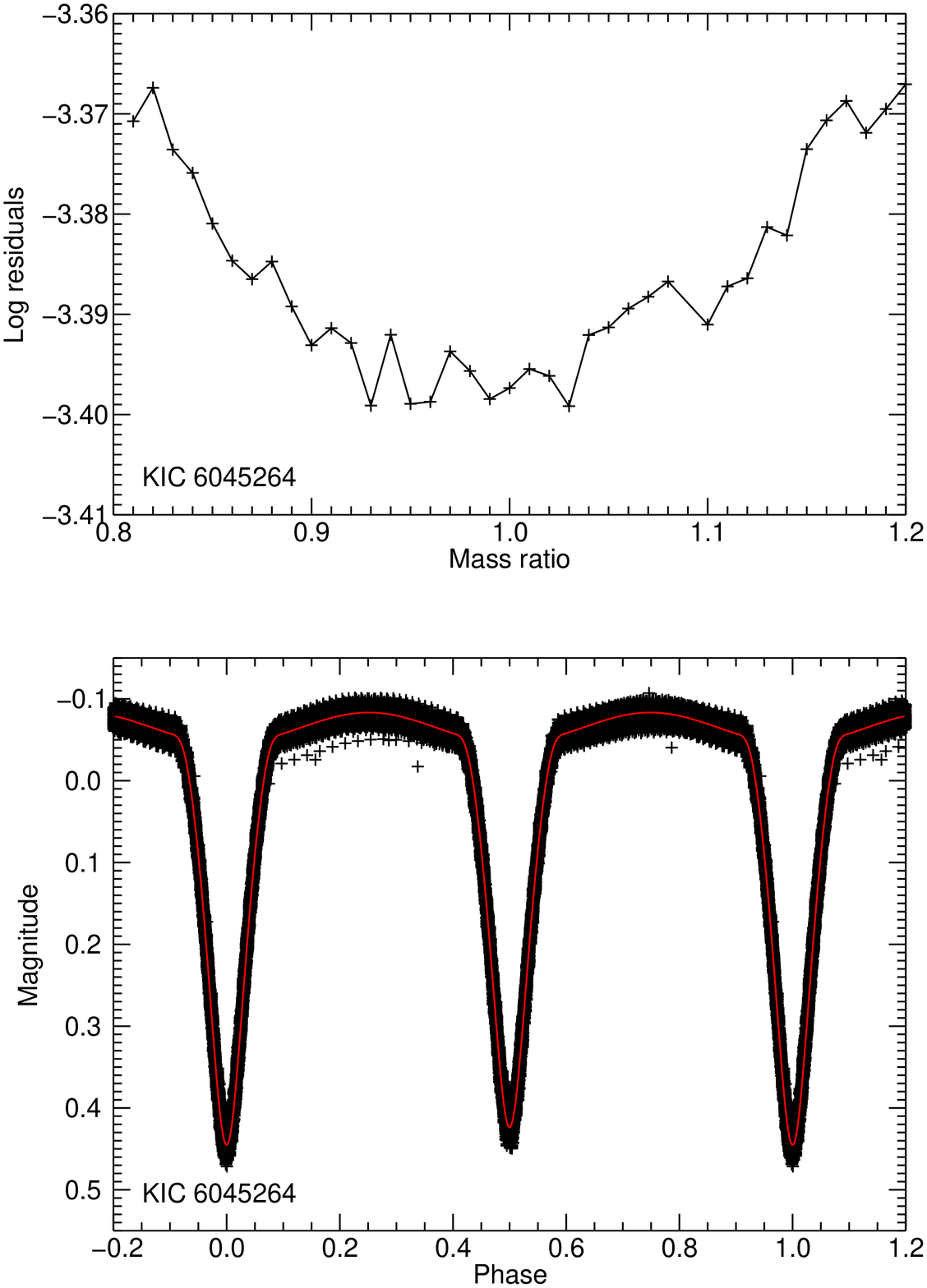}
\caption{Upper panels: the mass ratio search diagrams. Lower panels: All the observations folded to phase (black plus points) and the fits (red lines). The fitting curves were obtained based on the selected data, but also (should) fit all the observations well. The string of plus signs below the scatter band that extend from phases 0.1 to 0.3 in the lower right panel are some observations with large scatters, but they hardly have effect on the model fit due to their small quantity. \label{fig:wd}}
\end{minipage}
\end{figure*}

The fitting results are shown in Figure \ref{fig:wd} and listed in Table \ref{tab:wd}. In the final analysis, the independent variable of data is time, instead of phase. So the period, the zero point of the orbital ephemeris and the change rate of period can be derived directly by setting them to be adjustable. The resulting change rate of period is very small that can not be measured within current ability. The results show that, both for KIC 4826439 and KIC 6045264, the mass ratios, temperature ratios and radius ratios are all very close to unity, indicating that KIC 4826439 and KIC 6045264 are twin binaries.

\begin{small}
\begin{longtable}[h!]{lll}
\caption{The light curve solutions of KIC 4826439 and KIC 6045264.} \label{tab:wd}
  \hline
  \hline
Parameters & KIC 4826439 & KIC 6045264 \\
\endfirsthead
\endhead
  \hline
Mode                                                                      &    detached binary          &  detached binary       \\
Period (day)                                                              &    2.47429492(30)           & 0.90931038(16)         \\
Zero point of the orbital ephemeris $T_0$ (BJD)                           &    2454965.658114(38)       & 2454965.634999(56)     \\
Change rate of period $dP/dt$ (dimensionless)                             & $-0.037(\pm0.41)\times10^{-9}$ & $0.080(\pm0.22)\times10^{-9}$  \\
Orbital eccentricity $e$                                                  &    0 (assumed)           &  0 (assumed)    \\
Orbital inclination $i$ (degree)                                          &    88.2840(94)           &  83.339(80)     \\
Mass ratio $m_2/m_1$                                                      &    0.9909(41)            &  0.987(13)      \\
Primary temperature $T_1^{a}$ (K)                                         &    6065 (fixed)          &  6169 (fixed)   \\
Temperature ratio $T_2/T_1$                                               &    1.000824(94)          &  0.99157(24)    \\
Luminosity ratio $L_1/(L_1 + L_2)$ in band kepler                         &    0.49695(40)           &  0.4981(13)     \\
Luminosity ratio $L_2/(L_1 + L_2)$ in band kepler                         &    0.50305(40)           &  0.5019(13)     \\
Luminosity ratio $L_3/(L_1 + L_2 + L_3)$ in band kepler                   &    0.01558(43)           &  0.0204(35)     \\
Modified dimensionless surface potential of star 1                        &    6.4169(62)            &  4.971(14)      \\
Modified dimensionless surface potential of star 2                        &    6.358(20)             &  4.860(44)      \\
Radius of star 1 relative to semimajor axis in pole direction             &    0.18372(14)           &  0.24911(42)    \\
Radius of star 2 relative to semimajor axis in pole direction             &    0.18454(81)           &  0.2541(37)     \\
Radius of star 1 relative to semimajor axis in point direction            &    0.18693(15)           &  0.26132(53)    \\
Radius of star 2 relative to semimajor axis in point direction            &    0.18788(88)           &  0.2680(48)     \\
Radius of star 1 relative to semimajor axis in side direction             &    0.18486(14)           &  0.25306(44)    \\
Radius of star 2 relative to semimajor axis in side direction             &    0.18573(83)           &  0.2585(40)     \\
Radius of star 1 relative to semimajor axis in back direction             &    0.18645(15)           &  0.25862(49)    \\
Radius of star 2 relative to semimajor axis in back direction             &    0.18738(87)           &  0.2648(45)     \\
Equal-volume radius of star 1 relative to semimajor axis $R_1/A$          &    0.185096(75)          &  0.25383(24)    \\
Equal-volume radius of star 2 relative to semimajor axis $R_2/A$          &    0.18590(44)           &  0.2593(21)     \\
Radius ratio $R_2/R_1$                                                    &    1.0044(24)            &  1.0214(84)     \\
Roche Lobe equal-volume radius of star 1 relative to semimajor axis       &    0.37972(36)           &  0.3801(11)     \\
Roche Lobe equal-volume radius of star 2 relative to semimajor axis       &    0.37813(36)           &  0.3778(11)     \\
Filling degree of star 1 (the ratio of star volume to Roche Lobe volume)  &    0.11583(36)           &  0.2979(28)     \\
Filling degree of star 2 (the ratio of star volume to Roche Lobe volume)  &    0.11883(91)           &  0.3232(85)     \\
\hline
a: Obtained from the spectra analysis (see Section \ref{sec:The_spectra_analysis}) and fixed in the analysis.
\end{longtable}
\end{small}

\subsection{About the period and binarity of KIC 4826439}

Actually, before the light curves analysis, the period and binarity for KIC 4826439 should be clarified firstly. Since the primary and secondary eclipses are indistinguishable, one question arises: Is the object KIC 4826439 a circular orbit binary of 2.4742946 days period with identical primary and secondary eclipses, or a eccentric orbit binary with $2.4742946/2=1.2371473$ days period with only one eclipse? If the period is $2.4742946/2=1.2371473$ days, the phase length of the eclipse part is up to 0.2 large, indicating the two components are very close to each other. In such a close binary, it is almost impossible to have an eccentric orbit, and more impossible to have a large eccentricity to make the secondary eclipse disappear. So a very large eccentric orbit close binary system with only one eclipse is not possible for KIC 4826439. In addition, because the eclipse profile is wide and deep, it also can not be a planetary system. Therefore the object KIC 4826439 is a circular orbit binary with 2.4742946 days period. Especially, it can be perfectly fitted by a stellar binary model.

\section{The spectra analysis}\label{sec:The_spectra_analysis}

We adopt the ULySS (Universite de Lyon Spectroscopic analysis Software, \cite{2009A&A...501.1269K,2011A&A...525A..71W,2011RAA....11..924W,2014IAUS..306..340W}) package to analyze the spectra, it simultaneously determines the basic stellar atmospheric parameters (effective temperature $T_{eff}$, surface gravity $\log{g}$ and the metallicity $[Fe/H]$) via minimizing the $\chi^2$ between the observation and the model spectra by full spectra fitting, the fitting wavelength range is 4200-6800 {\AA}. The model spectrum is computed by an interpolator \citep{2011A&A...525A..71W} which is built based on the empirical ELODIE stellar library \citep{2001A&A...369.1048P,2007astro.ph..3658P}, it consists of a set of polynomial expansions of each wavelength element in powers of $T_{eff}$, $\log{g}$, $[Fe/H]$ and $f(\sigma)$ (a function of the rotational broadening parameterized by the velocity dispersion $\sigma$). Figure \ref{fig:spectra} illustrates the corresponding spectral fit.

The spectral data were measured as a single star, and therefore only one set of surface parameters were obtained. However, the two objects are all twin binaries with very close parameters between the components, thus the single set of parameters are valid for each component star. The results measured from three spectra are all considered, and the final results are listed in Table \ref{tab:spectra}.

\begin{longtable}[h!]{lll}
\caption{The atmospheric parameters of KIC 4826439 and KIC 6045264 derived from spectral data. \label{tab:spectra}}
\hline
  \hline
Parameters & KIC 4826439 & KIC 6045264 \\
\endfirsthead
\endhead
  \hline
$T_{eff}$ (K)               &    6065(45)      &  6169(25)     \\
$\log{g}$ ($cm/s^{2}$)      &    4.15(0.1)     &  4.18(0.1)     \\
$[Fe/H]^{a}$   &   -0.09(0.06)    &  -0.55(0.06)   \\
\hline
a: $[Fe/H]=\log_{10}(\frac{N_{Fe}}{N_{H}})_{star} - \log_{10}(\frac{N_{Fe}}{N_{H}})_{\odot}$
\end{longtable}

\section{The absolute parameters}

Generally, the radial velocity observations should be carried out to obtain the absolute parameters of a binary system. Here, we try to use another method to calculate the absolute parameters of the two objects KIC 4826439 and KIC 6045264. The light curves of the two objects were analyzed to obtain the relative parameters including the mass ratio $M_2/M_1$ or equivalently $M_{1,2}/M$ and the radius of stars relative to the semimajor axis $R_{1,2}/A$ (listed in Table \ref{tab:wd}), where $M$ and $A$ are the total mass and semimajor axis of a binary system respectively. On the other hand, the Kepler's law can give a relationship between the total mass $M$ and the semimajor axis $A$ since the orbital periods are known. Combined with the relative parameters $M_{1,2}/M$ and $R_{1,2}/A$ from the light curves analysis, the relationship between $M_1$ and $R_1$ (or $M_2$ and $R_2$) can be obtained. Therefore, the light curves analysis can provide a relationship between the stellar mass and radius. For the case of KIC 4826439 and KIC 6045264, the relationships of mass and radius for the primary stars are
\begin{eqnarray}
KIC 4826439: \qquad 5.761(\pm0.013)\frac{M_1}{M_\odot} = \left(  \frac{R_1}{R_\odot}  \right)^3, \nonumber \\
KIC 6045264: \qquad 2.003(\pm0.014)\frac{M_1}{M_\odot} = \left(  \frac{R_1}{R_\odot}  \right)^3. \label{equ:relationship}
\end{eqnarray}

\begin{figure*}[h!]
\begin{minipage}{160mm}
\centering
\includegraphics[width=.45\textwidth]{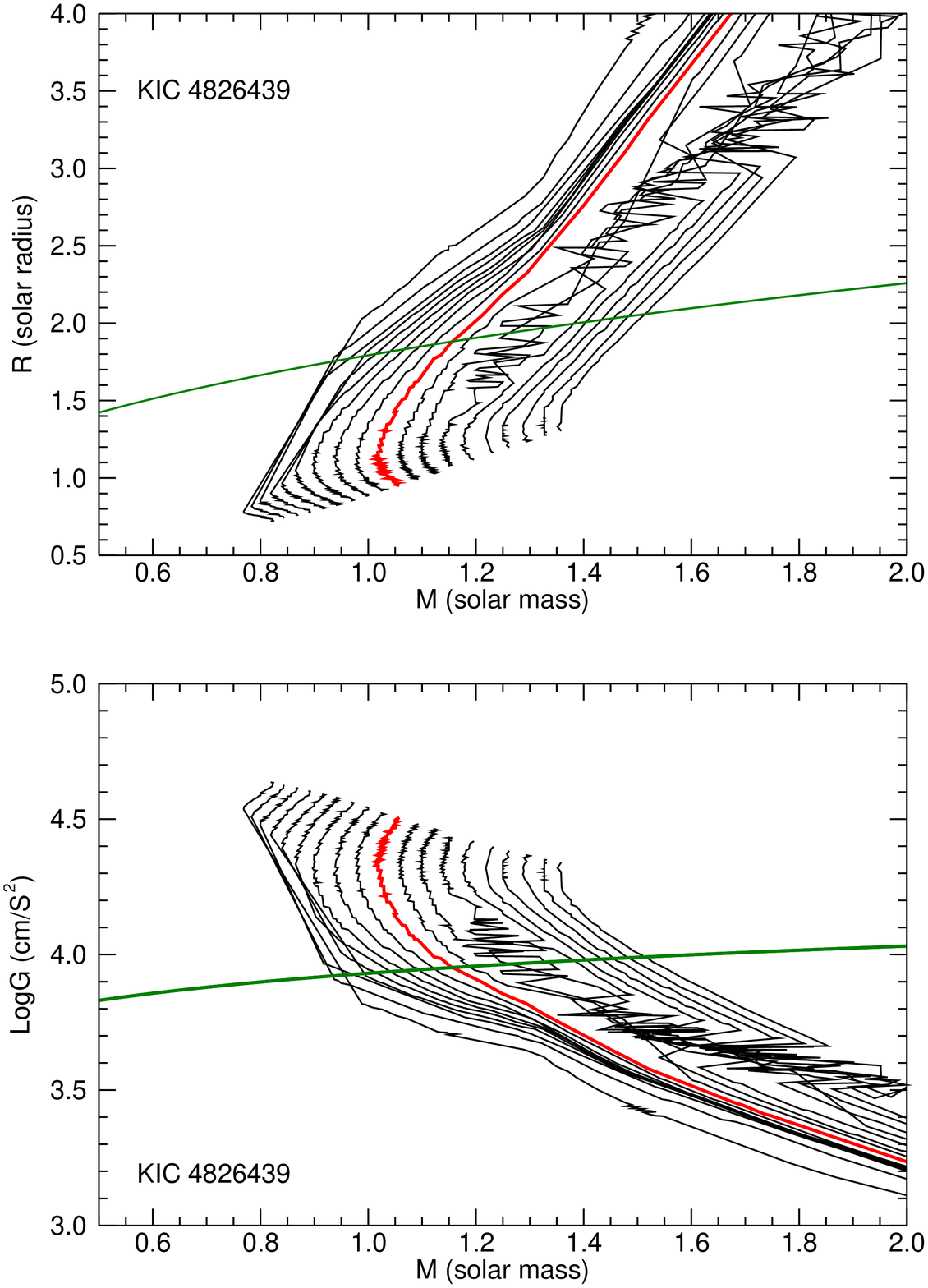}
\includegraphics[width=.45\textwidth]{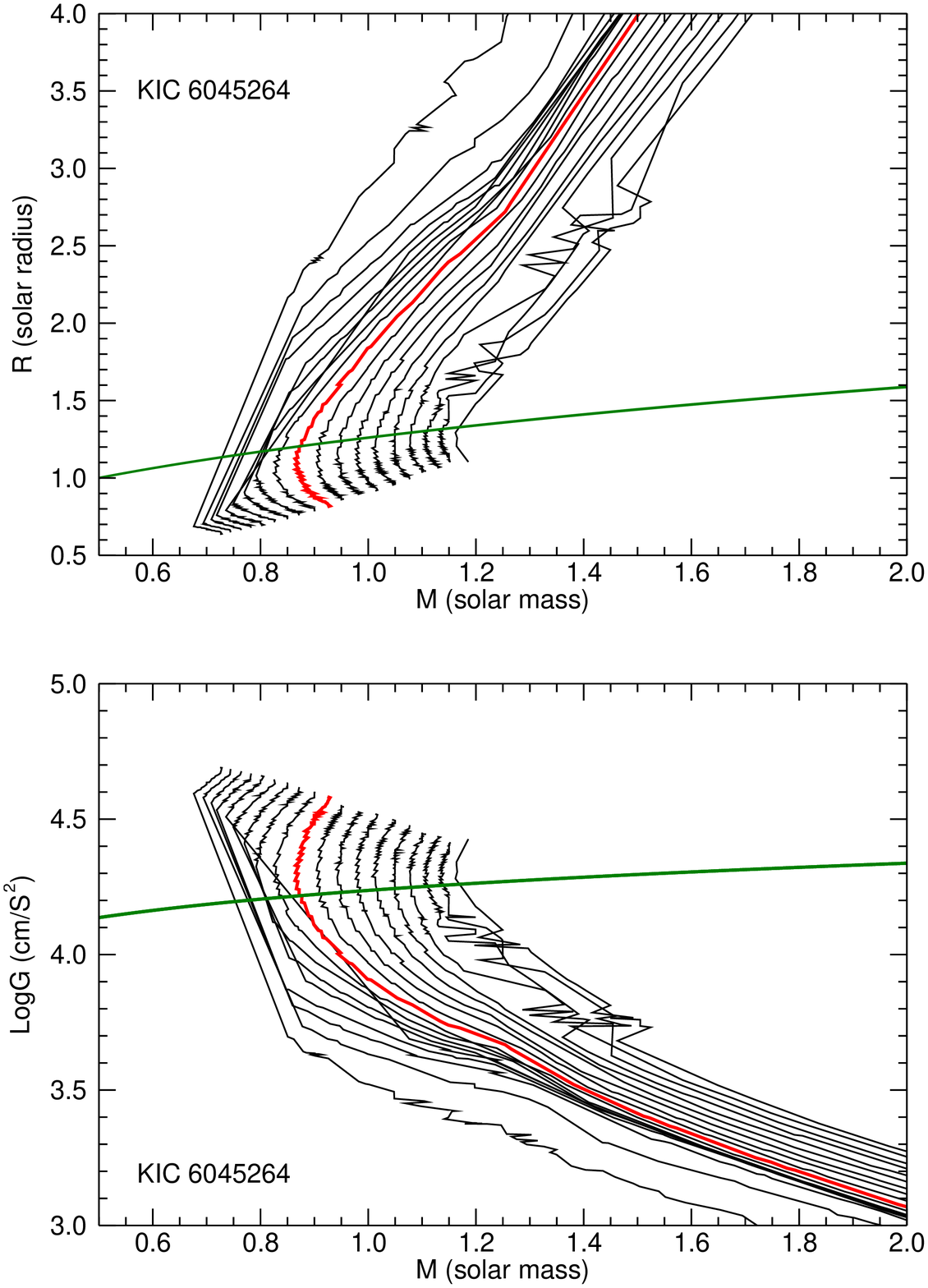}
\caption{The Mass-Radius (M-R) diagrams (upper panels) and the Mass-Surface gravity (M-LogG) diagrams (lower panels) for the primary components of KIC 4826439 (left panels) and KIC 6045264 (right panels). The black and red lines are the equal temperature lines, which are from 6065 - 900K to 6065 + 900K for KIC 4826439 and from 6169 - 900K to 6169 + 900 K for KIC 6045264 in a step of 100 K. The primary star temperatures, 6065 K (KIC 4826439) and 6169 K (KIC 6045264), are emphasized in the red thick lines. Some scrambled black lines indicates that the equal temperature regions may be wide. The green lines are the Mass-Radius relationships of the primary stars that described by Equation \ref{equ:relationship}.  \label{fig:MRG}}
\end{minipage}
\end{figure*}

In addition, the spectral data can provide another absolute parameter, i.e. the effective temperature (and with metallicity). In the real star models, the fixed temperature is also a limitation on other parameters. For example, in a mass-radius diagram the equal temperature zone is a line. The equal temperature lines in the Mass-Radius diagram are shown in the upper panels of Figure \ref{fig:MRG}. The isotherms have clear tracks without much chaos in the range of parameters displayed where the stellar masses are less than 5 $M_\odot$ and the temperatures ranged roughly from 5000 K to 7000 K. The red lines represent the temperatures measured from the spectra, and they have only one and clear intersection point with the green lines that represent the Mass-Radius relationships derived from the light curve analysis above.

The stellar model data used in Figure \ref{fig:MRG} are generated from the PAdova and TRieste Stellar Evolution Code (PARSEC) \citep{2012MNRAS.427..127B,2014MNRAS.444.2525C,2015MNRAS.452.1068C,2014MNRAS.445.4287T}, which can calculate evolutionary tracks for stars with various initial chemical compositions. A large mass range $0.1 \leq M/M_\odot < 350$ is investigated by the program to generate isochrones in a large age range $ 6.6 \leq \log(t/yr) \leq 10.12 $ in a step of 0.01. The lognormal form of initial mass function by \citet{2001ApJ...554.1274C} is used in the program, and the $\eta_{Remiers}$ is set to 0.2 in Reimers formula for mass-loss in RGB stage. The metallicity Z is set to $Z=0.012$ for KIC 4826439 and $Z=0.004$ for KIC 6045264.

The evolutionary code PARSEC covers almost all the stellar masses and ages, and then the equal temperature data picked up from them are the complete sample. The isotherms in the M-R diagram constitute a new Mass-Radius relationship (the red lines in Figure \ref{fig:MRG}) together with that from light curve analysis (the green lines in Figure \ref{fig:MRG}). These two relationships (or lines) match the sole mass and radius values for each object, and other parameters in the same way.

The temperature step between the adjacent isotherms is 100 K, but the isotherms are close to each other in M-R diagram. This imply that the errors on temperatures have little effects on the matching results. On the other hand, the errors of the green lines in Figure \ref{fig:MRG} are also very small, which sizes of the error bars are equal to the width of the lines themselves. Therefore, the parameters matched from the two relationship lines have small errors, which are listed in Table \ref{tab:abs_pars}.

\begin{longtable}[h!]{lll}
\caption{The absolute parameters of KIC 4826439 and KIC 6045264. \label{tab:abs_pars}}
\hline
  \hline
{Parameters} & {KIC 4826439} & {KIC 6045264} \\
\endfirsthead
\endhead
  \hline
%$Z$                &   0.012          &  0.004                \\
$M_1$ ($M_\odot$)  &  1.156(0.03)     &  0.874(0.03)          \\
$M_2$ ($M_\odot$)  &  1.145(0.03)     &  0.863(0.03)          \\
$R_1$ ($R_\odot$)  &  1.881(0.02)     &  1.206(0.02)          \\
$R_2$ ($R_\odot$)  &  1.889(0.02)     &  1.232(0.02)          \\
$T_1$ (K)          &  6065(45)        &  6169(30)             \\
$T_2$ (K)          &  6070(45)        &  6117(30)             \\
Age (years)      & $5.1(0.3)\times10^9$ & $9.6(0.3)\times10^9$   \\
Semimajor axix ($R_\odot$)  &  10.16(0.04)    &  4.75(0.03)       \\
Period   (day)        &  2.47429492(30)  &  0.90931038(16)        \\
\hline
\end{longtable}

The errors on temperature and metallicity were over taken into account for the errors in Table \ref{tab:abs_pars}. However, besides the atmosphere parameters, the input parameters of the evolution code also have influences on the outputs, and consequently on the final binary absolute parameters. The important input parameters include the helium abundance, the mixing-length parameter in the convection zone and the coefficient of Reimers mass-loss formula, which were set as $Y=0.2485+1.78Z$, $\alpha_{MLT}=1.74$ and $\eta_{Reimers}=0.2$, respectively. The reality of the evolution model and its input parameters heavily affected the reliability of the results in Table \ref{tab:abs_pars}. However, there is no extensive investigation on other values, which may lead to potential uncertainties on the results. The correctness of our method needs the verification from the radial velocity data.

In fact, the surface gravity measured from spectra is a more direct Mass-Radius relationship than the temperature. Thereby the mass and radius can be calculated directly from Equation \ref{equ:relationship} and the surface gravity without the evolutionary code. However the errors of the surface gravity have significant effects on the results, which can be seen from the lower panels of Figure \ref{fig:MRG}. The green lines are almost flat when entering the stellar mass range, and then a small change in surface gravity may lead to a large uncertainty in mass. According to the intersections in the lower panels of Figure \ref{fig:MRG}, the primary surface gravity $\log g_1$ is about 3.95 for KIC 4826439 and 4.22 for KIC 6045264, where the $\log g_1$ for KIC 4826439 is somewhat lower than that from the spectra analysis.

\section{the evolutionary tracks}

\begin{figure*}[h!]
\begin{minipage}{160mm}
\centering
\includegraphics[width=.45\textwidth]{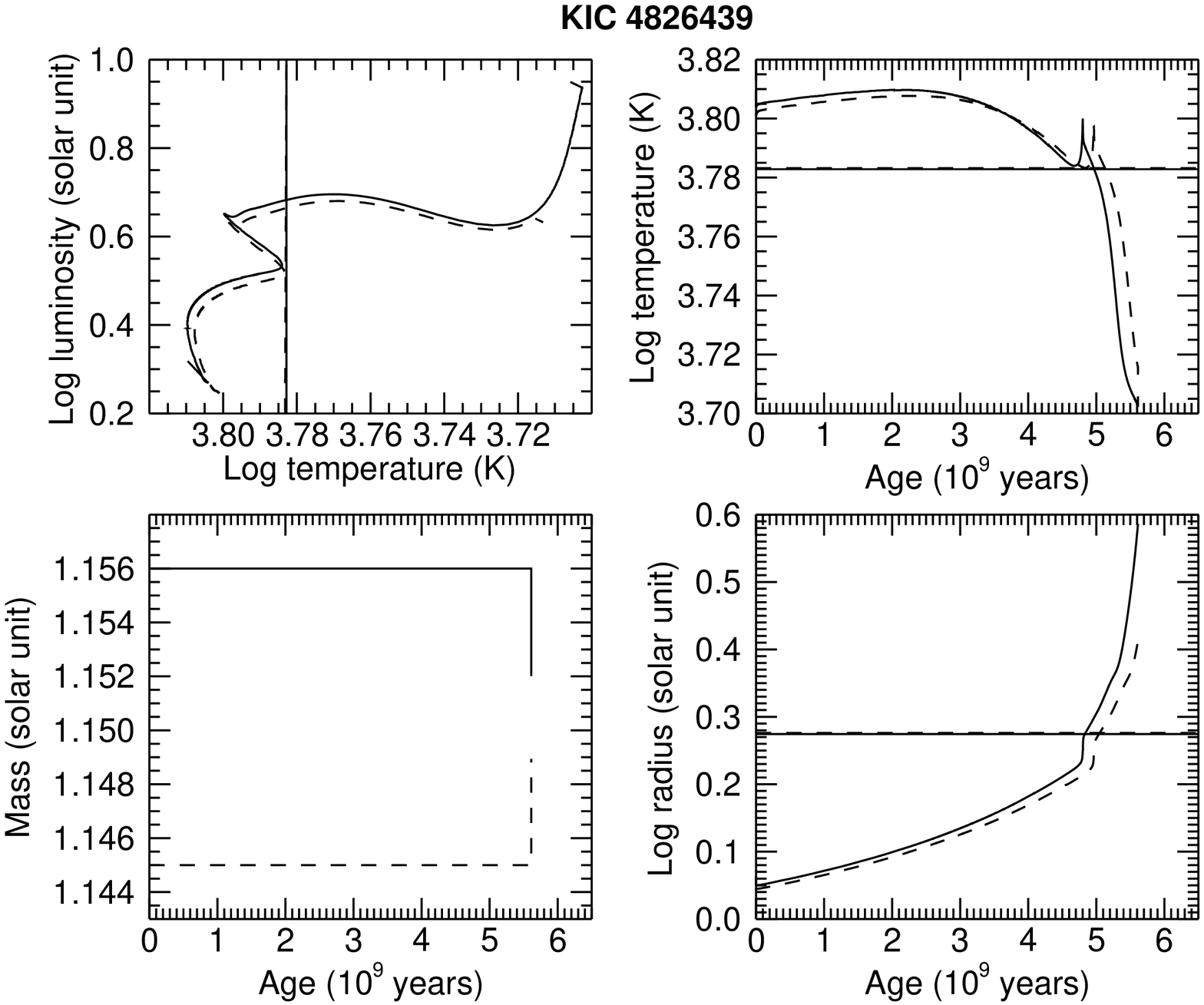}
\includegraphics[width=.45\textwidth]{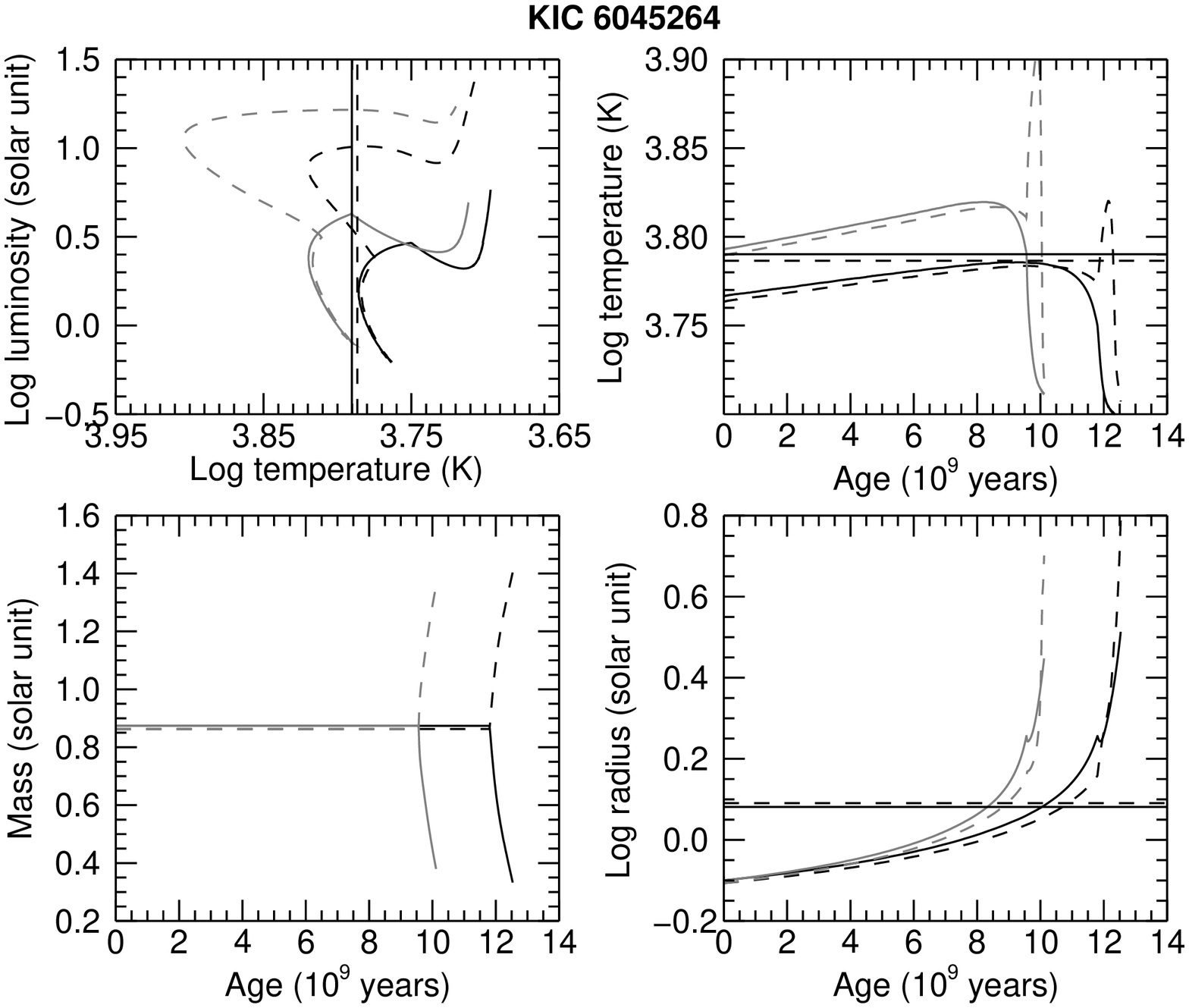}
\caption{The evolutionary tracks of KIC 4826439 (left panels) and KIC 6045264 (right panels). The solid and dashed lines stand for the massive and less massive component respectively, in which the vertical straight lines stand for the current parameter locations. The metallicity for KIC 4826439 is $Z=0.01$, and $Z=0.001$ (gray lines) and $Z=0.004$ (black lines) for KIC 6045264. \label{fig:evolution}}
\end{minipage}
\end{figure*}

It is wondered that will the twin binaries become twin degenerate binaries? So the evolutionary tracks of the two twins were calculated to study their endings. Here, the Eggleton's binary stellar evolution code was used which was originally proposed by \citet{1971MNRAS.151..351E,1972MNRAS.156..361E} and \citet{1973A&A....23..325E} and further developed by others, e.g. \cite{1994MNRAS.270..121H,1995MNRAS.274..964P,1998MNRAS.298..525P,2001ApJ...552..664N} and \citet{2002ApJ...575..461E}. The evolutionary tracks for the two twins were calculated and shown in Figure \ref{fig:evolution}.

There are two points concerning technical limitations need to be explained: 1, The masses and periods in Table \ref{tab:abs_pars} was used as the initial stellar parameters, and the mass loss by stellar wind was not considered in the evolution. This is because that if the stellar wind is taken into account, the initial period will be different from the present. However, it is difficult to guess the initial period. 2. Not all the metallicity within the error was calculated, but only three approximate metallicity, $Z=0.01$ for KIC 4826439 and $Z=0.001$ and $Z=0.004$ for KIC 6045264, were calculated. This is because the program does not allow arbitrary metallicity input due to the incomplete opacity table. Despite all this, the qualitative discussion on the evolutionary endings is still acceptable.

For KIC 4826439, the program stopped when two components just fill their Roche Lobe, where a very fast unstable mass transfer begins. It is generally believed that this kind of unstable mass transfer will lead to the ejection of the common envelope. It is still unclear that whether the final outcome of KIC 4826439 is a single degenerate or a binary system due to its moderate orbital period 0.9093104 days. For KIC 6045264, the program stopped when two components fill the outer Roche Lobe, where the system appearances a W UMa contact binary with mass ratio bigger than 4. It will finally merge into a fast rotating single star due to thermal instability \citep{1976ApJ...209..829W} or tidal instability \citep{1980A&A....92..167H,2010MNRAS.405.2485J,2013MNRAS.428.1218J}.

\section{summary and discussion} \label{sec:sum_dis}

Two twin binaries ($q>0.95$) were found photometrically from the \textit{Kepler} eclipsing binaries with period less than 5 days. Consequently, a very low proportion of twins $2/1592\approx0.13\%$ was found, which is significantly smaller the uniform distribution 5\%. This result does not support the previous findings that the twins proportion is excessively high on mass ratio distribution, but indicates a deficiently low proportion of twins. The absolute parameters of the two twins were obtained by a new method with the help of light curve analysis, atmosphere parameters and isochrones. Unfortunately, this method was not verified by the way of radial velocity, and so the high-resolution spectral observations are worthy to be carried out. The evolutionary endings of the two twins were also computed, and neither of them can produce a twin degenerate binary which indicates the formation conditions of twin degenerate binaries may be severe.

%%%%%%%%%%%%%%%%%%%%%%%%%%%%%%%%%%%%%%%

\begin{ack}
We thank Prof. Deng-Kai Jiang and Prof. Xiang-Cun Meng for their constructive works and suggestions that improve the manuscript significantly. New spectral observations were obtained with the Lijiang 2.4 m telescope at Yunnan Observatories.

Some of the data presented in this paper were obtained from the Mikulski Archive for Space Telescopes (MAST). STScI is operated by the Association of Universities for Research in Astronomy, Inc., under NASA contract NAS5-26555. Support for MAST for non-HST data is provided by the NASA Office of Space Science via grant NNX09AF08G and by other grants and contracts. This paper includes data collected by the \textit{Kepler} mission. Funding for the \textit{Kepler} mission is provided by the NASA Science Mission directorate.

This work is partly supported by the West Light Foundation of The Chinese Academy of Sciences, Yunnan Natural Science Foundation (Y5XB071001), Guangdong Provincial Engineering Technology Research Center for Data Science, the research fund of Sichuan University of Science and Engineering (grant No. 2015RC42), Chinese Natural Science Foundation (No. 11133007, 11325315, 11403056, 11573061, 11390374, 11373063 and U1631108), the Key Research Program of the Chinese Academy of Sciences (grant No. KGZD-EW-603), the Science Foundation of Yunnan Province (No. 2012HC011), and the Strategic Priority Research Program ``The Emergence of Cosmological Structures'' of the Chinese Academy of Sciences (No.XDB09010202).
\end{ack}

%\appendix
%\section*{The spectra of KIC 4826439 and KIC 6045264}
%\section{Case of two or paragraphs}
%
%\section{Case of two or paragraphs}

%%%
% See the manual for the detail.
%%%


\begin{thebibliography}{}
% Journals(e.g. A\&A,ApJ,AJ,NMRAS,PASP ...)
% Authors, Year, Journal, Vol#, Page#
% Journal Title Abbreviation >> http://www.asj.or.jp/pasj/Jabb.html
\bibitem[Armstrong et al.(2014)]{2014MNRAS.437.3473A} Armstrong, D.~J., G{\'o}mez Maqueo Chew, Y., Faedi, F., \& Pollacco, D.\ 2014, \mnras, 437, 3473
\bibitem[Bate(2000)]{2000MNRAS.314...33B} Bate, M.~R.\ 2000, \mnras, 314, 33
\bibitem[Borucki et al.(2010)]{2010Sci...327..977B} Borucki, W.~J., Koch, D., Basri, G., et al.\ 2010, Science, 327, 977   %Kepler_Mission
\bibitem[Bressan et al.(2012)]{2012MNRAS.427..127B} Bressan, A., Marigo, P., Girardi, L., et al.\ 2012, \mnras, 427, 127 %CMD2.8
\bibitem[Caldwell et al.(2010)]{2010ApJ...713L..92C} Caldwell, D.~A., Kolodziejczak, J.~J., Van Cleve, J.~E., et al.\ 2010, \apjl, 713, L92   %Kepler_Mission
\bibitem[Cantrell \& Dougan(2014)]{2014MNRAS.445.2028C} Cantrell, A.~G., \& Dougan, T.~J.\ 2014, \mnras, 445, 2028
\bibitem[Chabrier(2001)]{2001ApJ...554.1274C} Chabrier, G.\ 2001, \apj, 554, 1274
\bibitem[Chen et al.(2014)]{2014MNRAS.444.2525C} Chen, Y., Girardi, L., Bressan, A., et al.\ 2014, \mnras, 444, 2525  %CMD2.8
\bibitem[Chen et al.(2015)]{2015MNRAS.452.1068C} Chen, Y., Bressan, A., Girardi, L., et al.\ 2015, \mnras, 452, 1068  %CMD2.8
\bibitem[Conroy et al.(2014)]{2014AJ....147...45C} Conroy, K.~E., Pr{\v s}a, A., Stassun, K.~G., et al.\ 2014, \aj, 147, 45  %KEB4
\bibitem[Coughlin et al.(2014)]{2014AJ....147..119C} Coughlin, J.~L., Thompson, S.~E., Bryson, S.~T., et al.\ 2014, \aj, 147, 119
\bibitem[Eggleton \& Kiseleva-Eggleton(2002)]{2002ApJ...575..461E} Eggleton, P.~P., \& Kiseleva-Eggleton, L.\ 2002, \apj, 575, 461
\bibitem[Eggleton et al.(1973)]{1973A&A....23..325E} Eggleton, P.~P., Faulkner, J., \& Flannery, B.~P.\ 1973, \aap, 23, 325
\bibitem[Eggleton(1971)]{1971MNRAS.151..351E} Eggleton, P.~P.\ 1971, \mnras, 151, 351
\bibitem[Eggleton(1972)]{1972MNRAS.156..361E} Eggleton, P.~P.\ 1972, \mnras, 156, 361
\bibitem[Fraquelli et al.(2014)]{Fraquelli2014} Revision 5: D. Fraquelli \& S. E. Thompson, 2014, Kepler Archive Manual (KDMC-10008-005).
\bibitem[Gaidos \& Mann(2013)]{2013ApJ...762...41G} Gaidos, E., \& Mann, A.~W.\ 2013, \apj, 762, 41
\bibitem[Halbwachs et al.(2003)]{2003A&A...397..159H} Halbwachs, J.~L., Mayor, M., Udry, S., \& Arenou, F.\ 2003, \aap, 397, 159
\bibitem[Han et al.(1994)]{1994MNRAS.270..121H} Han, Z., Podsiadlowski, P., \& Eggleton, P.~P.\ 1994, \mnras, 270, 121
\bibitem[Hartman et al.(2004)]{2004AJ....128.1761H} Hartman, J.~D., Bakos, G., Stanek, K.~Z., \& Noyes, R.~W.\ 2004, \aj, 128, 1761
\bibitem[Hut(1980)]{1980A&A....92..167H} Hut, P.\ 1980, \aap, 92, 167
\bibitem[Jiang et al.(2010)]{2010MNRAS.405.2485J} Jiang, D., Han, Z., Wang, J., Jiang, T., \& Li, L.\ 2010, \mnras, 405, 2485
\bibitem[Jiang et al.(2013)]{2013MNRAS.428.1218J} Jiang, D., Han, Z., Yang, L., \& Li, L.\ 2013, \mnras, 428, 1218
\bibitem[Koch et al.(2010)]{2010ApJ...713L..79K} Koch, D.~G., Borucki, W.~J., Basri, G., et al.\ 2010, \apjl, 713, L79-L86   %Kepler_Mission
\bibitem[Koleva et al.(2009)]{2009A&A...501.1269K} Koleva, M., Prugniel, P., Bouchard, A., \& Wu, Y.\ 2009, \aap, 501, 1269
\bibitem[Krumholz \& Thompson(2007)]{2007ApJ...661.1034K} Krumholz, M.~R., \& Thompson, T.~A.\ 2007, \apj, 661, 1034
\bibitem[Krumholz et al.(2007)]{2007ApJ...656..959K} Krumholz, M.~R., Klein, R.~I., \& McKee, C.~F.\ 2007, \apj, 656, 959
\bibitem[Krumholz(2006)]{2006ApJ...641L..45K} Krumholz, M.~R.\ 2006, \apjl, 641, L45
\bibitem[Lucy \& Ricco(1979)]{1979AJ.....84..401L} Lucy, L.~B., \& Ricco, E.\ 1979, \aj, 84, 401
\bibitem[Lucy(2006)]{2006A&A...457..629L} Lucy, L.~B.\ 2006, \aap, 457, 629
\bibitem[Matijevi{\v c} et al.(2012)]{2012AJ....143..123M} Matijevi{\v c}, G., Pr{\v s}a, A., Orosz, J.~A., et al.\ 2012, \aj, 143, 123 %KEB3
\bibitem[Mulders et al.(2015)]{2015ApJ...814..130M} Mulders, G.~D., Pascucci, I., \& Apai, D.\ 2015, \apj, 814, 130
\bibitem[Mullally et al.(2015)]{2015ApJS..217...31M} Mullally, F., Coughlin, J.~L., Thompson, S.~E., et al.\ 2015, \apjs, 217, 31
\bibitem[Nelson \& Eggleton(2001)]{2001ApJ...552..664N} Nelson, C.~A., \& Eggleton, P.~P.\ 2001, \apj, 552, 664
\bibitem[Pigulski et al.(2009)]{2009AcA....59...33P} Pigulski, A., Pojma{\'n}ski, G., Pilecki, B., \& Szczygie{\l}, D.~M.\ 2009, AcA, 59, 33
\bibitem[Pinsonneault \& Stanek(2006)]{2006ApJ...639L..67P} Pinsonneault, M.~H., \& Stanek, K.~Z.\ 2006, \apjl, 639, L67
\bibitem[Pols et al.(1995)]{1995MNRAS.274..964P} Pols, O.~R., Tout, C.~A., Eggleton, P.~P., \& Han, Z.\ 1995, \mnras, 274, 964
\bibitem[Pols et al.(1998)]{1998MNRAS.298..525P} Pols, O.~R., Schr{\"o}der, K.-P., Hurley, J.~R., Tout, C.~A., \& Eggleton, P.~P.\ 1998, \mnras, 298, 525
\bibitem[Prugniel \& Soubiran(2001)]{2001A&A...369.1048P} Prugniel, P., \& Soubiran, C.\ 2001, \aap, 369, 1048
\bibitem[Prugniel et al.(2007)]{2007astro.ph..3658P} Prugniel, P., Soubiran, C., Koleva, M., \& Le Borgne, D.\ 2007, arXiv:astro-ph/0703658
\bibitem[Pr{\v s}a et al.(2011)]{2011AJ....141...83P} Pr{\v s}a, A., Batalha, N., Slawson, R.~W., et al.\ 2011, \aj, 141, 83 %KEB1
\bibitem[Sana et al.(2012)]{2012Sci...337..444S} Sana, H., de Mink, S.~E., de Koter, A., et al.\ 2012, Science, 337, 444
\bibitem[Simon \& Obbie(2009)]{2009AJ....137.3442S} Simon, M., \& Obbie, R.~C.\ 2009, \aj, 137, 3442
\bibitem[Slawson et al.(2011)]{2011AJ....142..160S} Slawson, R.~W., Pr{\v s}a, A., Welsh, W.~F., et al.\ 2011, \aj, 142, 160 %KEB2
\bibitem[Tang et al.(2014)]{2014MNRAS.445.4287T} Tang, J., Bressan, A., Rosenfield, P., et al.\ 2014, \mnras, 445, 4287  %CMD2.8
\bibitem[Tokovinin(2000)]{2000A&A...360..997T} Tokovinin, A.~A.\ 2000, \aap, 360, 997
\bibitem[Van Hamme (1993)]{1993AJ....106.2096V} van Hamme, W.\ 1993, \aj, 106, 2096 %limb darkening coefficient
\bibitem[Van Hamme \& Wilson(2007)]{2007ApJ...661.1129V} Van Hamme, W., \& Wilson, R.~E.\ 2007, \apj, 661, 1129  %WD
\bibitem[Webbink(1976)]{1976ApJ...209..829W} Webbink, R.~F.\ 1976, \apj, 209, 829
\bibitem[Wilson \& Devinney(1971)]{1971ApJ...166..605W} Wilson, R.~E., \& Devinney, E.~J.\ 1971, \apj, 166, 605  %WD
\bibitem[Wilson et al.(2010)]{2010ApJ...723.1469W} Wilson, R.~E., Van Hamme, W., \& Terrell, D.\ 2010, \apj, 723, 1469  %WD
\bibitem[Wilson(1979)]{1979ApJ...234.1054W} Wilson, R.~E.\ 1979, \apj, 234, 1054  %WD
\bibitem[Wilson(1990)]{1990ApJ...356..613W} Wilson, R.~E.\ 1990, \apj, 356, 613  %WD
\bibitem[Wilson(2008)]{2008ApJ...672..575W} Wilson, R.~E.\ 2008, \apj, 672, 575  %WD
\bibitem[Wilson(2012)]{2012AJ....144...73W} Wilson, R.~E.\ 2012, \aj, 144, 73  %WD
\bibitem[Wu et al.(2011a)]{2011A&A...525A..71W} Wu, Y., Singh, H.~P., Prugniel, P., Gupta, R., \& Koleva, M.\ 2011a, \aap, 525, A71
\bibitem[Wu et al.(2011b)]{2011RAA....11..924W} Wu, Y., Luo, A.-L., Li, H.-N., et al.\ 2011b, RAA, 11, 924
\bibitem[Wu et al.(2014)]{2014IAUS..306..340W} Wu, Y., Du, B., Luo, A., Zhao, Y., \& Yuan, H.\ 2014, IAUS, 306, 340
\end{thebibliography}
\end{document}